\documentclass[reprint,preprintnumbers,onecolumn,superscriptaddress,showkeys,
nofootinbib,notitlepage,amsmath,amssymb,floatfix]{revtex4-2}

\pdfoutput=1
\usepackage{latexsym,amsmath,amssymb,lmodern,float,url}
\usepackage{natbib}
\usepackage{bm}
\usepackage{mathtools}
\usepackage{xcolor}
\usepackage{comment}

\def\be{\begin{equation}}
\def\ee{\end{equation}}
\def\M{\mathcal{M}}

\def\I{\text{Im\,}}

\begin{document}
\preprint{JLAB-THY-26-4577}
\title{On Dispersive and Nondispersive K-matrix Formalisms}
\author{Nils H\"{u}sken}
\email{nhuesken@uni-mainz.de}
\affiliation{Johannes Gutenberg University of Mainz, Johann-Joachim-Becher-Weg 45, D-55099 Mainz, Germany}

\author{Eric~S.~Swanson}
\email{swansone@pitt.edu}
\affiliation{University of Pittsburgh, Pittsburgh, PA 15260, USA}

\author{Adam Szczepaniak}
\affiliation{Theory Center, Thomas  Jefferson  National  Accelerator  Facility,  Newport  News,  VA  23606,  USA}
\affiliation{Center for  Exploration  of  Energy  and  Matter,  Indiana  University,  Bloomington,  IN  47403,  USA}
\affiliation{Indiana University, Bloomington, Indiana 47405, USA}

\date{March, 2025}

\begin{abstract}

The modeling of coupled-channel effects has become increasingly important due to the availability of highly precise data for a large variety of hadronic (re)scattering processes. The K-matrix is a powerful, yet comparatively simple, method to describe scattering amplitudes, including coupled-channel effects, with the aim to interpret experimental data. Throughout the literature, a range of
dispersive and nondispersive K-matrix methods are employed.
Here, we compare the dispersive and nondispersive formulations in the context of the N/D method. It is shown that the methods are equivalent in the physical region under K-matrix reparameterization. Differences away from the physical region are examined.  Applications to synthetic data
are used to illustrate the effects of model choices concerning form factors and the application of dispersion relations, with the goal of clarifying best practices.
We find no clear preference with regards to dispersive modeling. In contrast, we find that interpretational ambiguity of the bare model parameters -- and even of the form of the bare model -- is endemic, and recommend a thorough sampling of data and model spaces to assess conclusion robustness.

\end{abstract}

\maketitle

\section{introduction}

The K-matrix formalism for scattering amplitudes in the presence of multiple open channels has a long history (see, e.g., Refs.~\cite{Dalitz:1960du,Aitchison:1972ay,Badalian:1981xj,Chung:1995dx} and references therein). It continues to be a useful tool in the analysis of experimental data for multiple reasons: by construction, an amplitude in K-matrix approach satisfies unitarity; in the case of two-particle scattering, the relation between the K-matrix and the transition matrix reduces to a simple algebraic equation; at the same time, the formalism is widely applicable due to its  flexibility. However, the K-matrix approach also comes with a number of drawbacks. Flexibility in model choices has lead to a wide range of different models being used throughout the literature with little consensus on whether some choices are ``better" or lead to equivalent results. In addition, the interpretation of the nature of resonances in terms of K-matrix parameters is unclear. For example Ref.~\cite{Chung:1995dx}, building on a discussion of Ref.~\cite{Au:1986vs}, offers two mechanisms to produce resonances: those arising from K-matrix pole terms, and those arising from constant K-matrix elements with energy variation introduced by the phase-space, relating the former to ``normal" resonances and the latter to molecular states. However, as we will relate, ambiguities present in K-matrix models raise the possibility that such interpretations are too simplistic.

Motivated by this lack of consensus, even for similar amplitude models, we have undertaken a systematic  examination of model-dependence in the unitary K-matrix approach to data fitting. In particular, we clarify the relationship between dispersive and nondispersive models, demonstrating that the methods are equivalent in the physical region. The difference away from the physical region is explored by examining a wide variety of fits to synthetic data. We find that neither method is generally preferred; thus computational ease and user preference should guide model choices. 

We have also examined the effect of form factors and dispersive subtraction constants on fit robustness. As hoped, the effect of these model choices can be largely absorbed by the model parameters, with the capability diminishing as one moves away from the physical region. Finally, we discuss the difficulty in  discerning spurious poles and the
commensurate lack of interpretability of  bare model parameters, or, indeed,  of the form of the amplitude model itself.

\section{amplitude modelling}

Our primary goal is to examine the robustness of  conclusions drawn from K-matrix models under a variety of conditions, mentioned in the Introduction. We thus give a brief review of the the K-matrix model, and then compare to the, more general, N/D method, which will be used to motivate the dispersive form of the K-matrix approach.  Of course, both formalisms are well-established, hence this section is primarily to set notation and introduce subsequent material. 

\subsection{The K-matrix Formalism}

As is well known, unitarity becomes an important constraint when resonances overlap. But allowing for the effects of coupled channels is perhaps more relevant to modern hadronic experiment. The K-matrix formalism is especially useful because it is unitary while permitting coupled-channel analyses in a simple way. We briefly outline the derivation of the K-matrix model here to set notation and to highlight an inherent ambiguity in the choice of phase space.

Unitarity of the S-matrix imposes a nonlinear relationship on the scattering amplitude that can be written in a generic form as

\be
\M - \M^\dagger = 2 i \M \rho \M^\dagger.
\label{eq:MM2}
\ee
The amplitude is defined in accordance with PDG conventions and is a matrix in coupled channel space, where channels are labeled according to particle content and all other relevant quantum numbers such as angular momentum, helicity, or reflectivity. These indices are collectively denoted with greek letters in the following. 
In the case of two particle scattering considered here, phase space is represented as a diagonal matrix given by,
\be
(\rho)_{\mu\gamma} \equiv \delta_{\mu\gamma}  \frac{k_\mu}{S_\mu 8 \pi \sqrt{s}}.
\label{eq:rho}
\ee
The  center of mass momentum for channel $\mu$ is given by
\be
k_\mu(s) = \sqrt{(s-s^\mu_0)(s-s^\mu_{ps})}/(2\sqrt{s}),
\ee
with $s^\mu_0 = (m^\mu_1+m^\mu_2)^2$ and $s^\mu_{ps} = (m^\mu_1-m^\mu_2)^2$ being the threshold and pseudothresholds for channel $\mu$.

Other conventions appear in the literature. The relative factor is absorbed into the reaction rates, yielding identical physical results. The convention adopted here is consistent with definitions given in the PDG for scattering [PDG section 49].

Direct manipulation of Eq. \ref{eq:MM2} yields 
\be
(\mathcal{M}^{-1}+i\rho)^\dagger = \mathcal{M}^{-1}+i\rho,
\ee
which implies that $\mathcal{M}^{-1}+i\rho$ is real. In the case that time reversal symmetry holds, this quantity is also symmetric.
It is thus traditional to define a real symmetric matrix $K$ via 
$\mathcal{M}^{-1} + i \rho = K^{-1}$. However, motivated by the comparison to quantum field theory, it can be convenient to combine a portion of the K-matrix with the phase space. Thus a complex diagonal matrix $C$ is introduced via 
\be
\mathcal{M}^{-1} = K^{-1} + C,
\label{eq:Kdefn} 
\ee
with $\I\,C  = - \rho$, and 
where, of course, the diagonal portion of $K$ has shifted from the previous expression.
With this definition 
\begin{equation}
\M = (1+KC)^{-1}K = K(1+CK)^{-1} = K(K+KCK)^{-1}K.
\label{eq:M}
\end{equation}

K-matrix models generally assume that an amplitude is dominated by s-channel resonance exchange. 
In analogy to scalar field theory, it is common to write a K-matrix ansatz in the form

\be
K_{\mu\nu} = \sum_R \frac{g_{R:\mu} g_{R:\nu}}{m_R^2-s} + f_{\mu:\nu}(s)
\label{eq:kk}
\ee
where the sum is over bare poles. The additional term represents nonresonant ``background" scattering. 
Unitarity implies that the residue of the amplitude must factorize at the location of a real pole (see Section 6.1 of \cite{Martin:1970hmp}. This is generalized to complex poles in Ref. \cite{Gribov:2009zz}, page 84). 
Furthermore, it is known that amplitudes must behave as $k_\mu^{\ell_\mu}\cdot k_\nu^{\ell_\nu}$  where $\ell_\mu$ is the leading partial wave in channel $\mu$ (see Section 8.1 of Ref. \cite{Martin:1970hmp}). Thus it is common to model the couplings as form factors $g_{R:\mu} \to g_{R:\mu}(s)$ and to set $f_{\mu:\nu} = g_\mu(s) \hat f_{\mu:\nu}(s) g_\nu(s)$, with 

\be
g_{R:\mu}(s) \to k_\mu^{\ell_\mu} 
\label{eq:thresh}
\ee
near threshold.

The presence of these form factors is  interpreted as being due to the soft nature of hadronic interactions. They are often included because the rapid growth of $k_\mu^{\ell_\mu}$ for large $s$ is regarded as unphysical. They are also required by the  positivity postulates of quantum mechanics. Specifically, the K\"{a}llen-Lehmann representation for the full two point function implies that a scalar propagator $\Delta(s)$ with unit residue must behave as $1/s$ (more precisely as $s^n$ for $n\geq -1$) for large $|s|$, thus $\rho(s)g^2(s)$ can diverge at most like $s$ (see Section 10.7 of  Ref. \cite{Weinberg:1995mt}). 

In a similar fashion, choosing the background scattering model is something of a black art, with typical models relying on intuition or simplicity considerations. 

The remaining issue is determining the form of the matrix $C$. 
A common choice is to set its real part to zero. While fulfilling unitarity, this has the unfortunate consequence of retaining nonanalyticities in the phase space at $s=0$ and $s=s_{ps}$, and therefore tends to be deprecated as a model choice. A way forward is provided by considerations of analyticity, to which we turn.

\subsection{The N/D Formalism}

The  $N/D$ formalism was originally introduced by Chew and Mandelstam\ in a description of $\pi\pi$ scattering~\cite{Chew:1960iv}, where the authors noted that the integral over a phase shift like
\begin{equation} 
D(s) = \exp\left( -\frac{1}{\pi}\int ds' \cdot \frac{\delta(s')}{s'-s-i\epsilon}\right)
\end{equation} 
is analytic except for a right hand cut and has the phase $\exp(-i\delta)$ in the physical region. Thus the amplitude is proportional to $D^{-1}(s)$ and must have a numerator function that is real in the physical region and that only contains possible left hand cuts. 
Thus one defines $\M = N/D$. In the physical region $\I D = N \,\I\M^{-1}  = - \rho N.
$
Similarly, on the left hand cut
$ \I N = D \,\I\M
$ (indices are removed in this section to reduce notational bloat).
Thus
 
\be
D(s) = \sum_{i=0}^{m-1} c_i s^i + \sum_j \frac{d_j}{s_j - s}  - \frac{s^m}{\pi} \int_{s_0} ds' \frac{\rho(s') N(s')}{s^{'m}(s'-s-i\epsilon)},
\label{eq:Ddisp}
\ee
where $m$ subtractions have been made at the point $s=0$. The second term represents possible Castillejo-Dalitz-Dyson (CDD) poles that can appear in $D$, whose properties are independent of the constraints imposed by unitarity. The presence of a CDD  pole implies a zero in the scattering amplitude, which implies a nearby pole. Thus CDD poles are often associated with quark model bound states. In a similar fashion, the dispersion relation for the numerator function is 
\be
N(s) = n(s) + \frac{s^p}{\pi} \int_{-\infty}^{s_L} ds' \frac{D(s') \, \I M(s')}{s^{'p}(s'-s-i\epsilon)}.
\label{eq:Ndisp}
\ee

Two benefits are apparent with this approach: left hand cut physics can be accommodated and the poor analytic behavior of the phase space and the form factors has been ameliorated by placing these under the dispersion integral of Eq. \ref{eq:Ddisp}. For example, the singularities at $s=0$ and $s=s_{ps}$ in $\rho$ are not encountered in the dispersion integral because they lie below threshold. The dispersion relation also resolves ambiguities in defining the breakup momentum below threshold, again, because $k(s)$ always appears within dispersion integrals. 

The physics associated with left hand cuts is not considered in K-matrix models and the expression for $N(s)$ can be simplified by  neglecting the integral in Eq. \ref{eq:Ndisp}. Thus the most general unitary amplitude  with no left hand cuts can be written as 
\be
\M^{-1} = {\mathcal R}(s) - \int_{s_0} \frac{ds'}{\pi} \frac{s^m}{{s'}^m}\frac{n(s')}{n(s)}\frac{\rho(s')}{(s'-s-i\epsilon)}
\label{eq:KD}
\ee
where ${\mathcal R}(s)$ is a rational function of $s$ that is real and has no left hand or right hand cuts. Alternatively, the identification of the polynomial  term in  Eq.~\ref{eq:Ddisp} as an order $(m-1)$ pole at $s\to \infty$ reveals that $R$ subsumes all CDD poles.

It is convenient to choose the function $n$ to reflect the known threshold behavior of the system because this brings the singularities in the breakup momentum under the dispersion integral.  Extending this slightly, one can set $n = g^2(s)$ to mimic the K-matrix model. 
Comparison with Eq. \ref{eq:Kdefn} identifies $\mathcal{R} = K^{-1}$ and 

\be
C = - \int_{s_0} \frac{ds'}{\pi} \frac{g^2(s')}{g^2(s)}\, \frac{\rho(s')}{s'-s-i\epsilon},
\label{eq:KD2}
\ee
where $m$ has been set to zero because the dispersion integral no longer needs subtraction.

At this stage it is useful to define quantities $\mathcal{M} = g \hat{\mathcal{M}}g$ and $K = g \hat K g$ so that the K-matrix equation becomes
\be
\hat{\mathcal{M}}^{-1} = \hat K^{-1} - \int_{s_0} \frac{ds'}{\pi}\, \frac{\rho(s') g^2(s')}{s'-s-i\epsilon}.
\label{eq:KD3}
\ee
This is useful because all the dependence on the breakup momentum  is shifted to phase space (in $C$). Of course, if one starts with the $N$ over $D$ representation of the scattering amplitude,  then $K$ is given by the left hand cuts and CDD poles.  
This definition of $C$ serves to extend its validity away from the physical region in a way that is consistent with analyticity, thereby achieving the goal of ameliorating unphysical singularities in the phase space. 

Of course,  the full amplitude contains powers of $g$, 
so that poor behavior in the coupling models still appears in the amplitude. Thus it is unclear whether employing Eq. \ref{eq:KD3} or Eq. \ref{eq:KD} with the simple choice $n=1$ (which we call the semidispersive form) is the preferred approach. This will be examined in depth in Section \ref{iii}.

\subsection{Relationship of the nondispersive and dispersive forms of the K-matrix}

We examine the relationship between the dispersive and semidispersive forms of the K-matrix. Specifically, we aim to demonstrate that they are equivalent in the physical region with sufficiently general K-matrix parameterizations, which means establishing that
$$
\hat K_0^{-1} -  \int \frac{ds'}{\pi} \frac{\rho(s')}{s'-s-i\epsilon} = \hat K_1^{-1} - \int \frac{ds'}{\pi}\frac{g^2(s')}{g^2(s)}\frac{\rho(s')}{s'-s-i\epsilon}
$$
where $K_0$ and $K_1$ must both be real in the physical region. The generalization to the coupled channel case presents no obstacles. Indeed, the equivalence follows immediately because the imaginary part of the difference between the left and right sides is

\be
\I \delta C(s) \equiv \I \int \frac{ds'}{\pi}\left(\frac{g^2(s')}{g^2(s)}-1\right)\frac{\rho(s')}{s'-s-i\epsilon} = 0
\label{eq:dC}
\ee
when $s$ is real. 
    Thus, one can simply set

\be
\hat K_0^{-1} = \hat K_1^{-1} - \delta C \label{eq:dC2}
\ee
to obtain a valid equivalent K-matrix. 
We note that it is also possible that 
$\delta C$  has singularities away from the physical region.

As an example consider P-wave Blatt-Weisskopf form factors in the equal mass case. Then 

\be
g^2(s) = \frac{k^2}{\beta^2+k^2} = \frac{s-s_0}{s-s_0+s_B}
\ee
with $s_B=4\beta^2$. In this case the integral is logarithmically divergent. If the term proportional to $\rho(s')$ is regulated with a subtraction, $c_0$ and the term proportional to $g^2(s')\rho(s')$ is regulated with a subtraction $c_1$, then

\be
\delta C = 
\frac{c_1}{g^2(s)} - c_0 + \frac{\tanh^{-1}(y)}{8 \pi^2 y}, \ \  y = \sqrt{1-s_0/s_B}.
\ee

If $c_1=0$ and  $\hat K_1^{-1}$ is parameterized as a polynomial then obtaining $\hat K_0^{-1}$ is trivial. Alternatively, 
if one employs the Ansatz of Eq. \ref{eq:kk}, then the explicit mapping from $\hat K_1$ to $\hat K_0$ can be made if one defines a new bare pole at the solution to $m_R^2-s+g^2 \delta\mathcal{C}=0$, say, $s={M_R'}^2$. Expanding around this minimum and defining new coupling and background models gives

\be
{g'}^{2}(s) = g^2(s)\left[1-\frac{d}{ds}(g^2\delta\mathcal{C})|_{s={M_R'}^{2}}\right]^{-1}
\ee
and
\be
f'(s) = -\frac{1}{2}{g'}^{2}(s)\, \frac{d^2}{ds^2}(g^2\delta\mathcal{C})|_{s={M_R'}^{2}}\,\left[1-\frac{d}{ds}(g^2\delta\mathcal{C})|_{s={M_R'}^{2}}\right]^{-1} + O({M_R'}^{2}-s).
\ee
Notice that the coupling model is simply scaled by a factor, which can be absorbed into the bare couplings of the K-matrix; furthermore, if the scattering is parameterized as $f_{\mu\nu}(s) = \hat f_{\mu\nu}\cdot g_\mu(s)g_\nu(s)$ then all changes  up to $O({M_R'}^2-s)$ can be accommodated by rescaling bare couplings, $\hat f_{\mu\nu}$ and $\hat g_\mu$.

\subsection{Sheet Structure}

The primary output of any K-matrix analysis is the location of resonance poles and their residues. Obtaining these poles relies on extending the amplitude model into the complex plane, which requires knowledge of the analytic structure of the K-matrix. Unfortunately, 
information concerning the analytic structure of the K-matrix is scattered and sometimes inconsistent, we therefore discuss it here in some detail.

First, it is traditional to make cuts due to threshold openings move to the right. This can be implemented by setting 

\be
k_\mu \to k_{RHC} = \pm i \sqrt{(s_\mu-s)(s-s_{ps})}/(2\sqrt{s}).
\ee
Recall that $s$ should always be interpreted as $s+i\epsilon$ if not stated explicitly. 

The sheet structure of an amplitude determines the proximity of poles to the physical region, with nearby poles dominating the behavior of the reaction. Sheets are  distinguished by the signs of the breakup momenta, $\{sgn(k_\mu)\}$. We order these according to energy to assign a unique identifier to each sheet. Thus the physical sheet, traditionally called sheet I, is \mbox{$\langle +++\ldots\rangle$}. As first discussed in Ref. \cite{Frazer:1964zz}, crossing the real axis above the first
threshold connects this to sheet \mbox{$\langle-++\ldots\rangle$} (``sheet II"), which is the closest resonance sheet between the first and second thresholds. Crossing the second threshold moves to sheet \mbox{$\langle --+\ldots\rangle$} (``sheet III"). Similarly, in this region sheet \mbox{$\langle +-+\dots\rangle$} connects to \mbox{$\langle -++\ldots\rangle$} (sheet IV to sheet II). As a result, the relevant sheet for a resonance above the $n$th threshold is \mbox{$\langle(-\ldots)^n+++\ldots\rangle$}.

This exercise is important because poles on distant sheets have little to do with measured reactions and because they are increasingly likely (as we will see in Section \ref{iii}) to be artifacts of the model, rather than physical features. 

Amplitudes are constrained to be continuous as one moves across the real axis, thus the discontinuity across the cut must be added to the amplitude as it crosses from one sheet to another. Thus, for example,

\be
\lim_{\epsilon\to 0}\mathcal{M}_I(s+i\epsilon)|_{PR} = \lim_{\epsilon\to 0}\mathcal{M}_{II}(s-i\epsilon)|_{PR}.
\ee
For the factored form of the K-matrix formalism (Eq. \ref{eq:KD3}), this discontinuity is given by 

\begin{eqnarray}
C_{\mu\mu} \to\ && C_{\mu\mu} - 2 i \rho_\mu^*(k_{RHC}(s^*)) \, g^{2*}_\mu(k_{RHC}(s^*)) \nonumber \\ 
            =\ &&  C_{\mu\mu}+2 i \rho_\mu(k_{RHC}(s))\, g_\mu^{2}(k_{RHC}(s))\nonumber \\
            =\ &&  C_{\mu\mu} - 2i \rho_\mu(k(s))\, g_\mu^2(k(s)).
\end{eqnarray} 
We stress that changing the sign of $k_\mu$ or forming the complex conjugate of $C$ is \textit{not} adequate for defining the sheet structure of the amplitude.

Lastly, we remark that $k_\mu$ is imaginary for $ s_{ps} < s < s_\mu$, which appears to do damage to the required real nature of the K-matrix if factors of the breakup momentum appear in the couplings. This problem is, however, obviated if couplings are grouped with phase space, as in the factored approach.

\section{Applications and Model-dependence}
\label{iii}

\begin{figure}[ht]
  \centering
  \includegraphics[width=0.45\columnwidth]{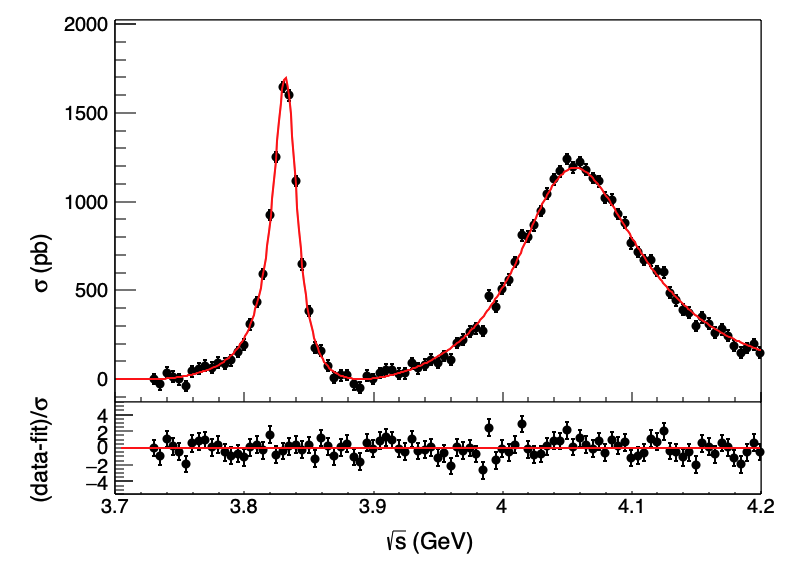}
  \qquad
\includegraphics[width=0.45\columnwidth]
{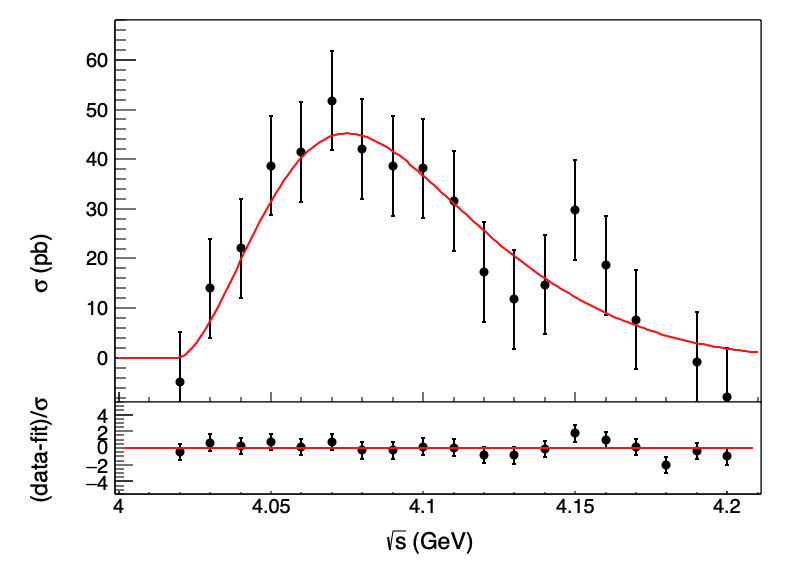}
  \caption{$D_0\bar D_0$ (left) and $D^*\bar D^*$ data and fit (same model).}
  \label{fig:D0D0-fit0}
\end{figure}

Several issues need to be addressed. These include selecting the  phase factor $C$ and understanding the effects of different form factor models. We aim to explore these issues by analyzing synthetic data produced from a simple two-channel model designed to mimic the  production of $e^+e^-\to D\bar D$ and $D^*\bar D^*$  pairs. The data are generated using a dispersive K-matrix model with p-wave Blatt-Weisskopf form factors with an interaction radius of $r_0 = 1/\mbox{GeV}$, and incorporating two bare poles (ie, poles in the K-matrix), and Gaussian noise (see Fig. \ref{fig:D0D0-fit0}). The resulting physical poles are located at

$$
P_1 = 3833 - 12i \ (\text{MeV}) \ \ P_2 = 4049 -64 i\ \text{(MeV)},
$$
which are near the bare poles and are  well-separated. Variations of this model are considered in Section \ref{sect:checks} to confirm that our conclusions are not overly dependent on the dataset.  

A variety of models will be used to fit these data; these are comprised of various combinations of 

i) phase space: $C=-i\rho$, semi-dispersive, and dispersive models;

ii) form factors: $k^\ell\, \exp(-k^2/\beta^2)$, Blatt-Weisskopf, and $k^\ell/(k^2+\beta^2)$ are considered;

iii) K-matrix models with 0, 1, or 2 bare poles;

iv) form factor scale $\beta = 0.2 - 4.0$ GeV in steps of $0.2$ GeV;

v) the inclusion of subtraction constants used (or not) as fit parameters;

and

vi) the degree of the background scattering polynomial in $s$ is varied from zero to two.

\subsection{Moving Off-axis}

\begin{figure}[ht]
  \centering
  \includegraphics[width=0.5\columnwidth]{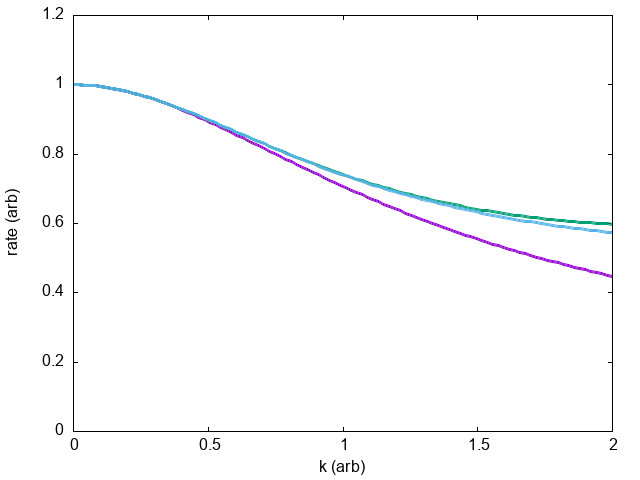}
  \caption{Three Amplitude Models in the Physical Region.}
  \label{fig:1}
\end{figure}

\begin{figure}[ht]
  \centering
  \includegraphics[width=0.3\columnwidth]{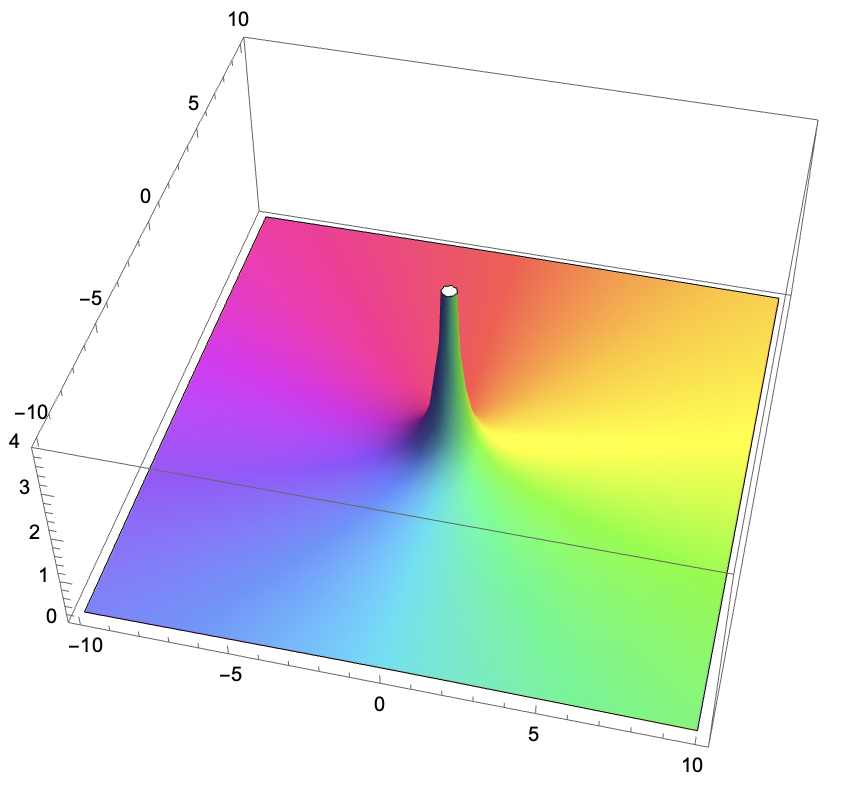}
   \includegraphics[width=0.3\columnwidth]{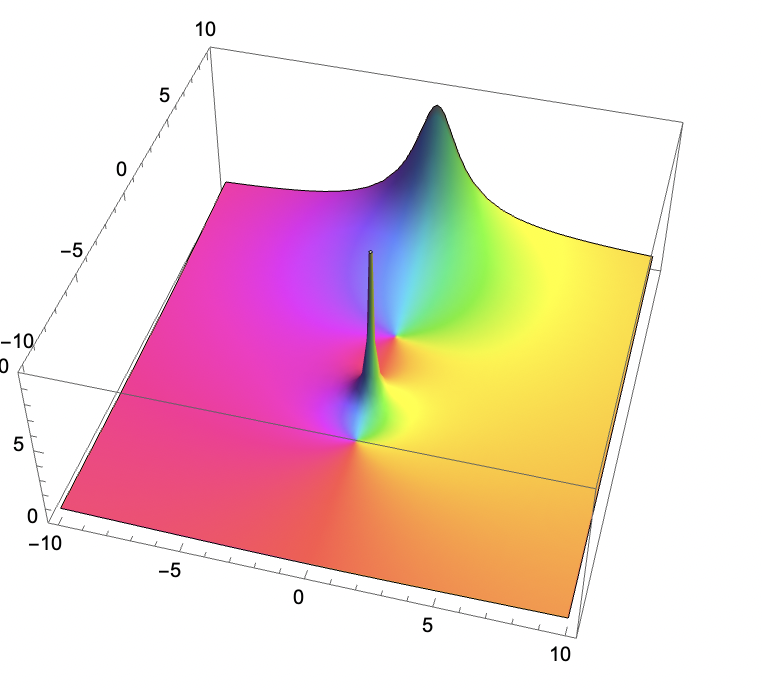}
    \includegraphics[width=0.3\columnwidth]{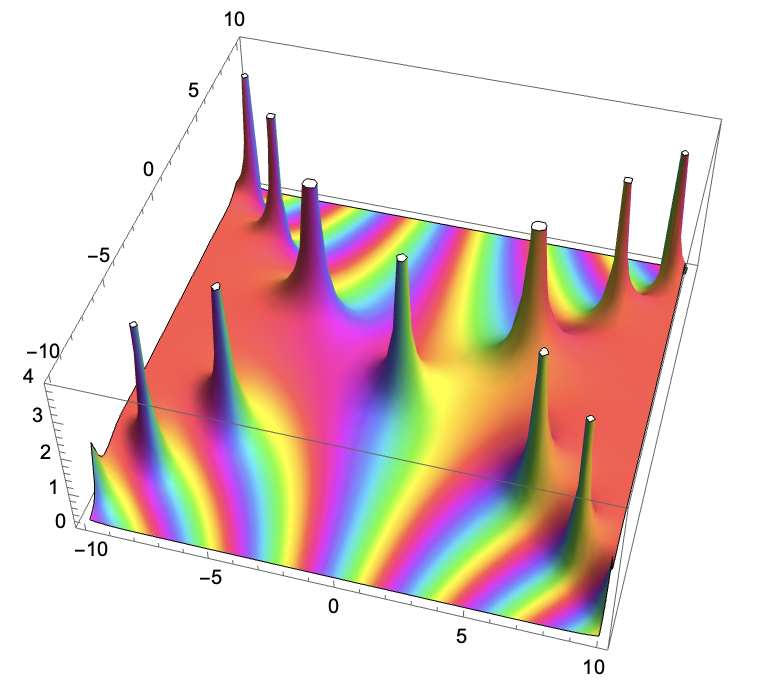}
  \caption{Three Amplitude Models in the Complex Plane. From left to right these map to the curves of Fig. \ref{fig:1} from  bottom to top.}
  \label{fig:2}
\end{figure}

In a restricted energy range, the differences between common model choices can be mostly absorbed by model parameters, so similar accuracy is expected when fitting data. Likewise, functions that are almost identical on the real axis should also be nearly the same when continued into the nearby complex plane. This holds up to a scale where behavior can differ significantly. This is illustrated in Figs. \ref{fig:1} and \ref{fig:2}, which show three simple complex functions designed to emulate a simple pole, a model with a Blatt-Weisskopf form factor, and a model with an exponential form factor. As seen in Fig. \ref{fig:1}, the three functions look very similar in the ``physical region." The situation changes in the complex plane, where extra structure appears once one moves away from the pole location. In particular, the exponential model exhibits an infinite set of poles, spaced roughly on the order of the exponential scale from the root pole shown in the leftmost plot.  This simple example shows that extraneous poles related to form factors can be easily generated, so exploring a range of amplitude models is important. Furthermore, poles become less reliable the farther they are from the physical region. This will be quantified to some degree in the following.

\subsection{Bare Parameters are not Physical}

If a good fit is found, there is a natural tendency to see the fit model as ``truth." For example, a pole unrelated to bare poles in the model will be viewed as a dynamical resonance. As this section's examples will demonstrate, this is incorrect. The situation is similar to quantum field theory, where bare parameters have no physical meaning, as their dependence on the renormalization scale shows. In fact, the model itself has no physical meaning—we will see that models with two, one, or no bare poles fit the synthetic data well, clearly illustrating that overinterpreting bare poles (or models) can be risky.

\begin{figure}[ht]
  \centering
  \includegraphics[width=0.45\columnwidth]{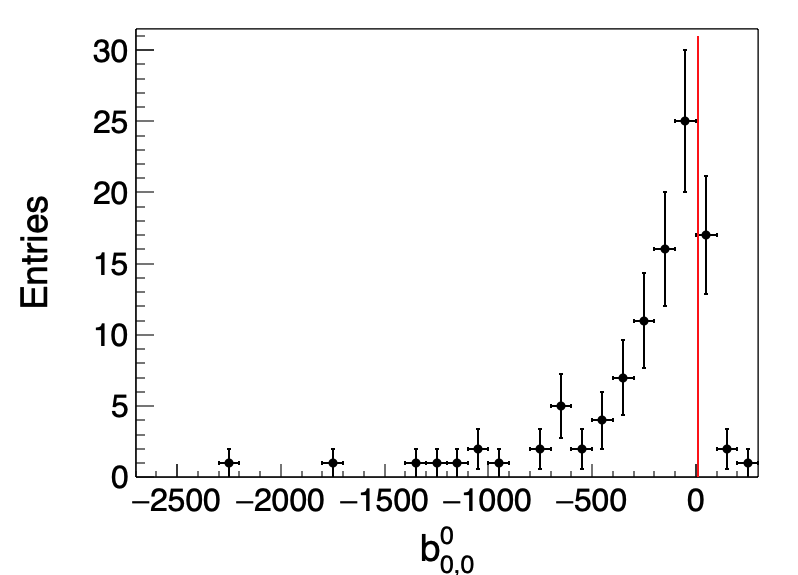}
  \qquad
    \includegraphics[width=0.45\columnwidth]{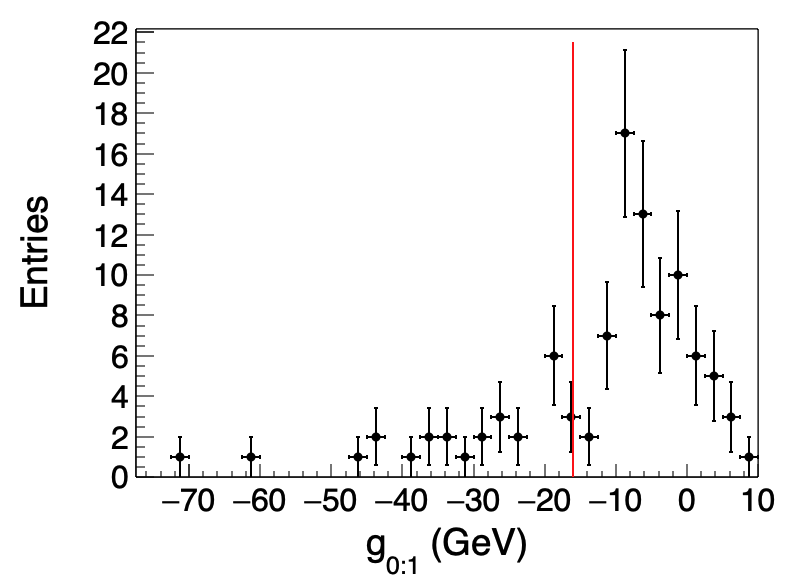}
    \caption{Distribution of $D \bar D$ background term (left) and $R1: D^*\bar D^*$ coupling (right).}
  \label{fig:iocheck_b00}
\end{figure}

\begin{figure}[ht]
  \centering
  \includegraphics[width=0.5\columnwidth]{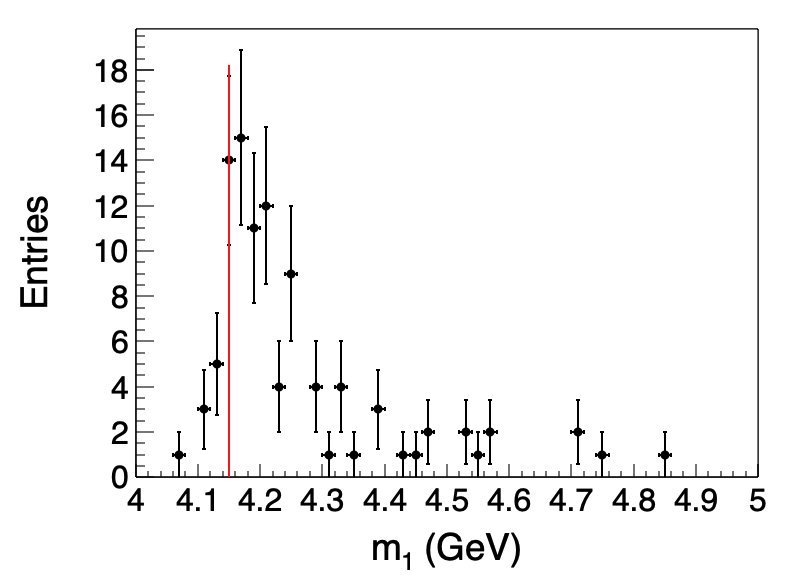}
    \caption{Distribution of second bare mass.}
  \label{fig:iocheck_m1}
\end{figure}

We start by fitting the synthetic data to the model that generated it. Naturally, this should perform well. In fact, the model is flexible enough to accommodate some fluctuations in the data, resulting in $\chi^2$/dof = 0.95. The result is shown as the solid line in Fig. \ref{fig:D0D0-fit0}. The resulting pole locations are unsurprisingly stable.
$$
P_1 = 3833 - 12i\ \text{MeV}, \ \ P_2 = 4048 -64i \ \text{MeV}.
$$
In addition, the couplings to the production channel, $g_{R:ee}$ are accurately reproduced. However, resonance couplings experience some drift, $g_{R1:\alpha} = (10,-16)$ changes to fit values of  $(9.2,-15.7)$ and  
$g_{R2:\alpha} = (15,6)$ changes to $(16.7, 2.8)$.
In contrast to this stability, the model background parameters are \textit{not} stable. To better quantify this, we generated 100 bootstrap samples from the dataset and fitted these with the generating model. The frequency histograms for selected model parameters are shown in Figs. \ref{fig:iocheck_b00} and \ref{fig:iocheck_m1}. It is evident, even in this ideal scenario, that the spread of parameters is quite broad. Therefore, data fluctuations alone can produce widely varying model parameters. Clearly, it is incorrect to ascribe reality or over-interpret parameters when their intrinsic variability can be so significant.

The challenge in interpreting model parameters also applies to the models themselves. To demonstrate this, we examined the fit efficacy of a wide variety of models, as detailed above.  
The dispersion integral of Eq. \ref{eq:KD3}  is known for the P-wave Blatt-Weisskopf case with equal mass particles~\cite{Hanhart:2023fud}). In other cases it is computed numerically on a fine grid, and precomputed values are interpolated during fitting.

Models with $\chi^2 > 125$ are not shown in the following figures. The number of parameters ranges from 9 to 17, so the maximum acceptable reduced chi-squared is $ \chi^2/dof=1.29$ (with a corresponding p-value of 0.03). A total of 392 models were fitted using different combinations of assumptions, and 320 of these met the plotting criteria. For each successful fit, 100 bootstrap samples of synthetic data were generated to assess the statistical uncertainty of the pole properties relative to the systematic variation between different models.

\begin{figure}[ht]
  \centering
  \includegraphics[width=0.45\columnwidth]{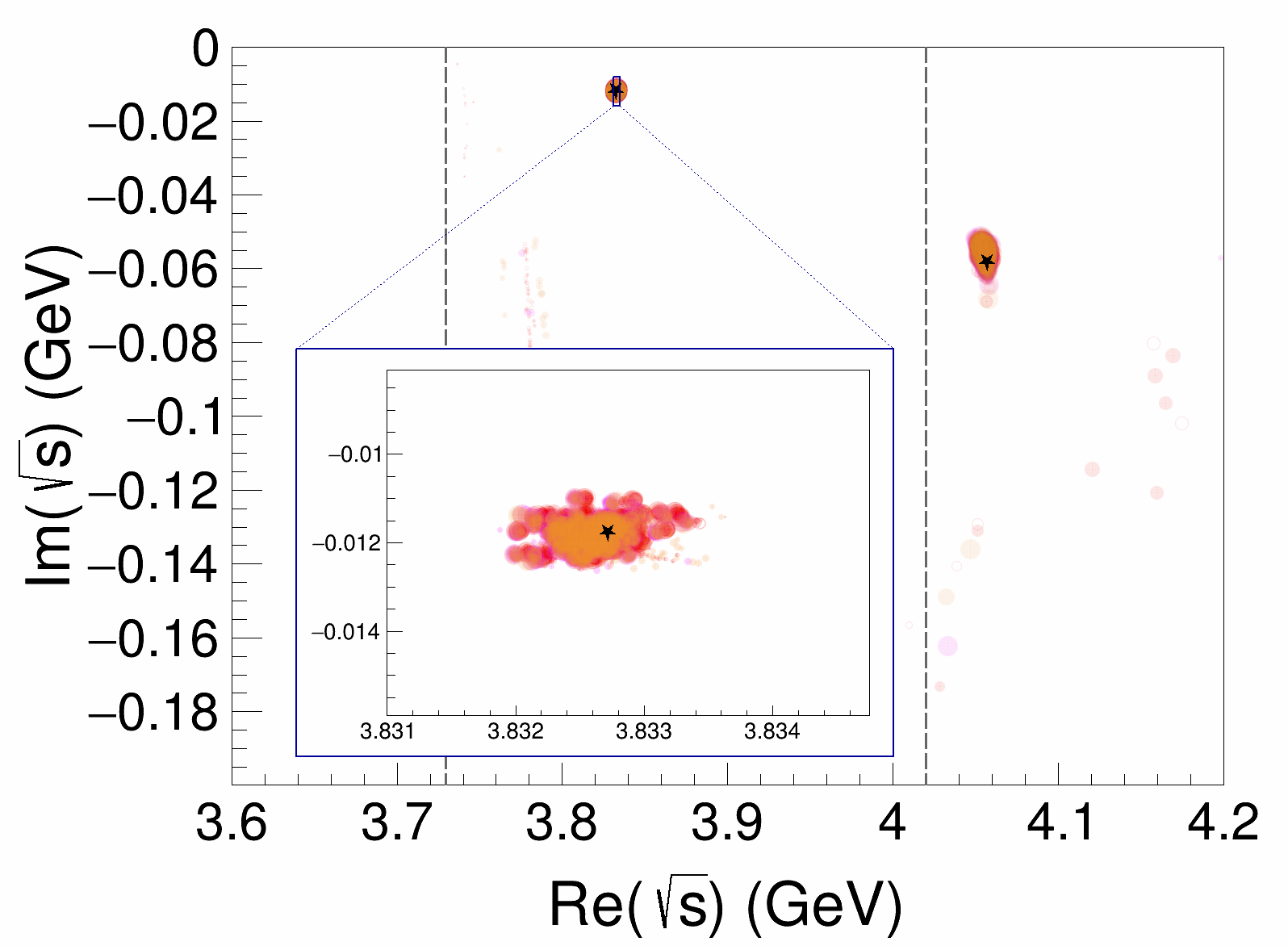}
  \qquad
    \includegraphics[width=0.45\columnwidth]{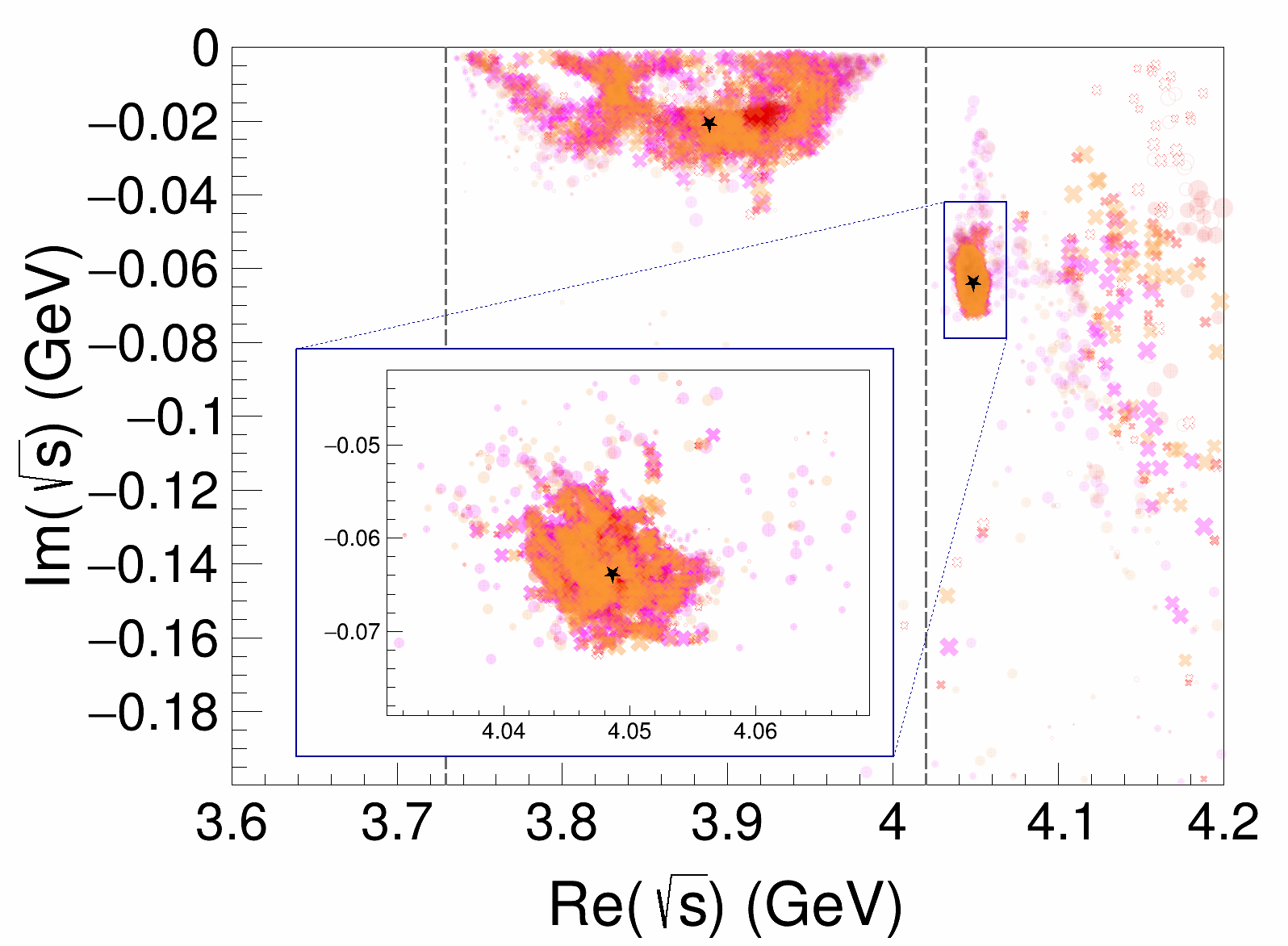}
    \caption{Pole positions for a variety of dispersive models with two bare poles. Blatt-Weisskopf (red points); power form factor (orange); exponential (pink).  Models with subtraction constants as parameters are indicated with open symbols. Symbol sizes are proportional to $\beta$.  Crosses indicate solutions with more than  two poles.
    Sheet $\langle -+\rangle$ (II) (left); sheet $\langle --\rangle$ (III) (right).}
  \label{fig:2pole}
\end{figure}

The resulting pole positions are shown in Fig. \ref{fig:2pole} for all dispersive models with two bare poles and constant background terms, so the only variation of the input is the coupling model. A tight group of poles centered on the nominal pole appears for the sheet $\langle-+\rangle$ (II). A similar cluster is visible on the sheet $\langle --\rangle$ (III), but a much larger spread is observed in the pole positions, especially for the lower mass pole, for which this sheet is distant from the physical region. These unstable poles appear across all categories of models and all values of form factor scales considered.  

Crosses indicate model fits with extra poles. Evidently, the collection of poles on sheet $\langle--\rangle$ starting to the right of the nominal pole is dominated by solutions with spurious poles, while solutions with two poles tend to cluster more around the nominal pole.

It is also worth noting that the lowest poles are more scattered on sheet $\langle--\rangle$ than on sheet $\langle-+\rangle$; indicating that the increased distance from the physical region is causing additional deviations in the pole locations--an effect almost certainly due to energy dependence in the model couplings. Finally, we observe that the choice of form factor has little impact on pole positions, except that spurious poles tend to be more common when fitting with model form factors not used to generate the synthetic data, namely the exponential and power law models.

\begin{figure}[ht]
  \centering
  \includegraphics[width=0.45\columnwidth]{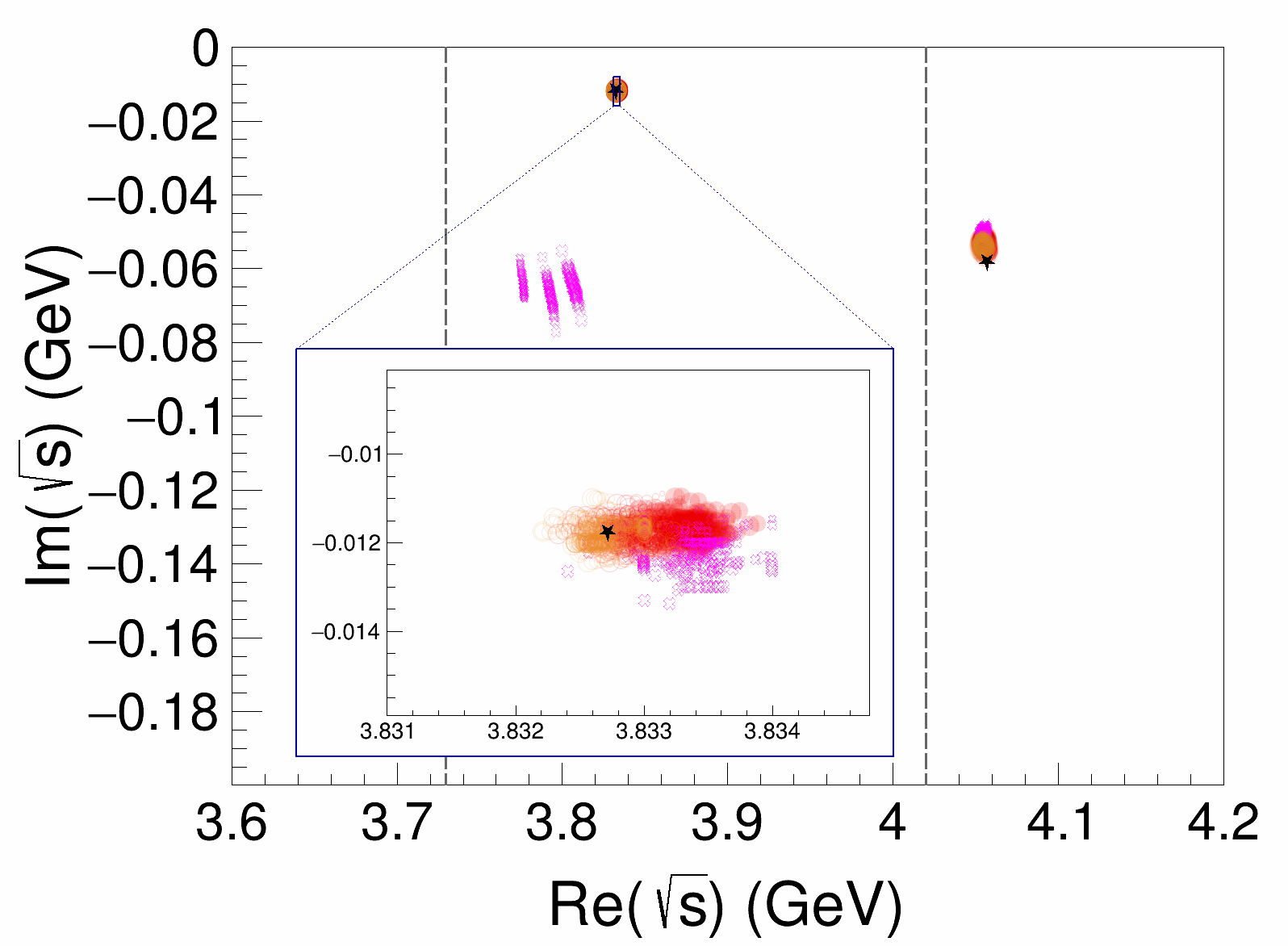}
  \qquad
    \includegraphics[width=0.45\columnwidth]{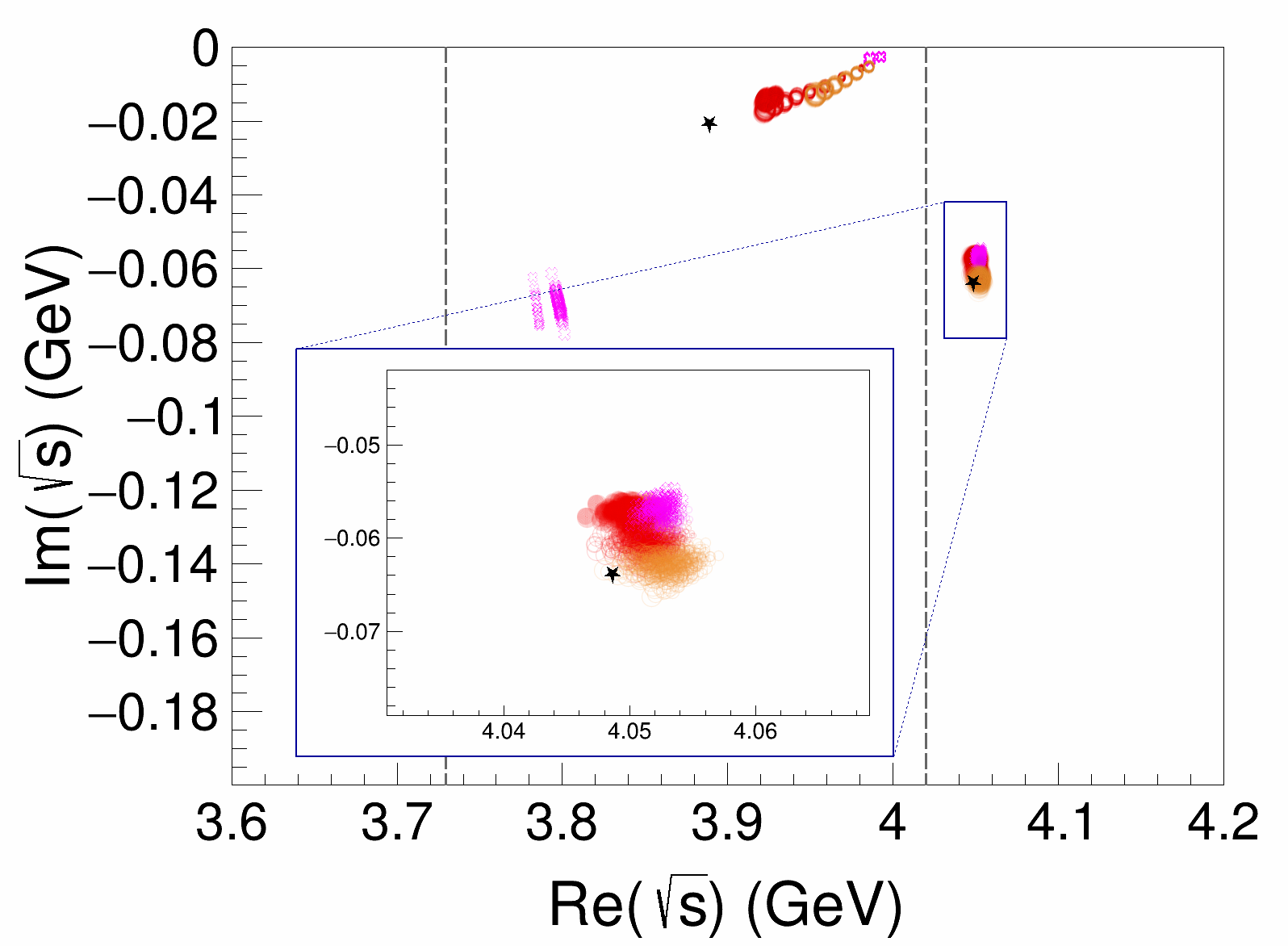}
    \caption{One bare pole. Symbols are as in Fig. \ref{fig:2pole}.}
  \label{fig:1pole}
\end{figure}

The exercise was repeated with the same model variations but using a single bare pole, producing the results shown in Fig. \ref{fig:1pole}. Fit quality was similar to that of the two-pole models, so it cannot serve as a metric to distinguish between the classes of models. Once again, the nominal poles are accurately reproduced; however, interestingly, the poles on sheet $\langle--\rangle$ tend to be slightly heavy in mass and narrow in width.
Additionally, there is a sequence of extraneous poles on sheet $\langle-+\rangle$ that are produced by models with extra subtraction constant parameters. Clearly, these poles are spurious (in the sense that they are not stable and not related to the nominal poles); evidently the additional degree of freedom allows a good fit to the data while creating an extra pole.

Notably, the first pole differs considerably from its nominal position on the third sheet (right panel), likely due to its greater distance from the physical region. This becomes more apparent in the relationship.

\be
\mathcal{M}_{\text{IV}}^{-1} = \mathcal{M}_{\text{III}}^{-1} - 2 i \rho_{\text{III}} g^2.
\ee
Since resonances are defined by $\text{det}\mathcal{M}^{-1}=0$, we observe that poles on the two sheets are closely connected if the phase space factor varies smoothly. Thus poles that that are closely related to a bare pole tend to have images on related sheets\cite{Au:1986vs}. However, if this factor changes rapidly or if one is in a regime where the form factor model becomes unreliable, then poles on higher sheets can differ significantly from those on their physically relevant sheets.

\begin{figure}[ht]
  \centering
  \includegraphics[width=0.45\columnwidth]{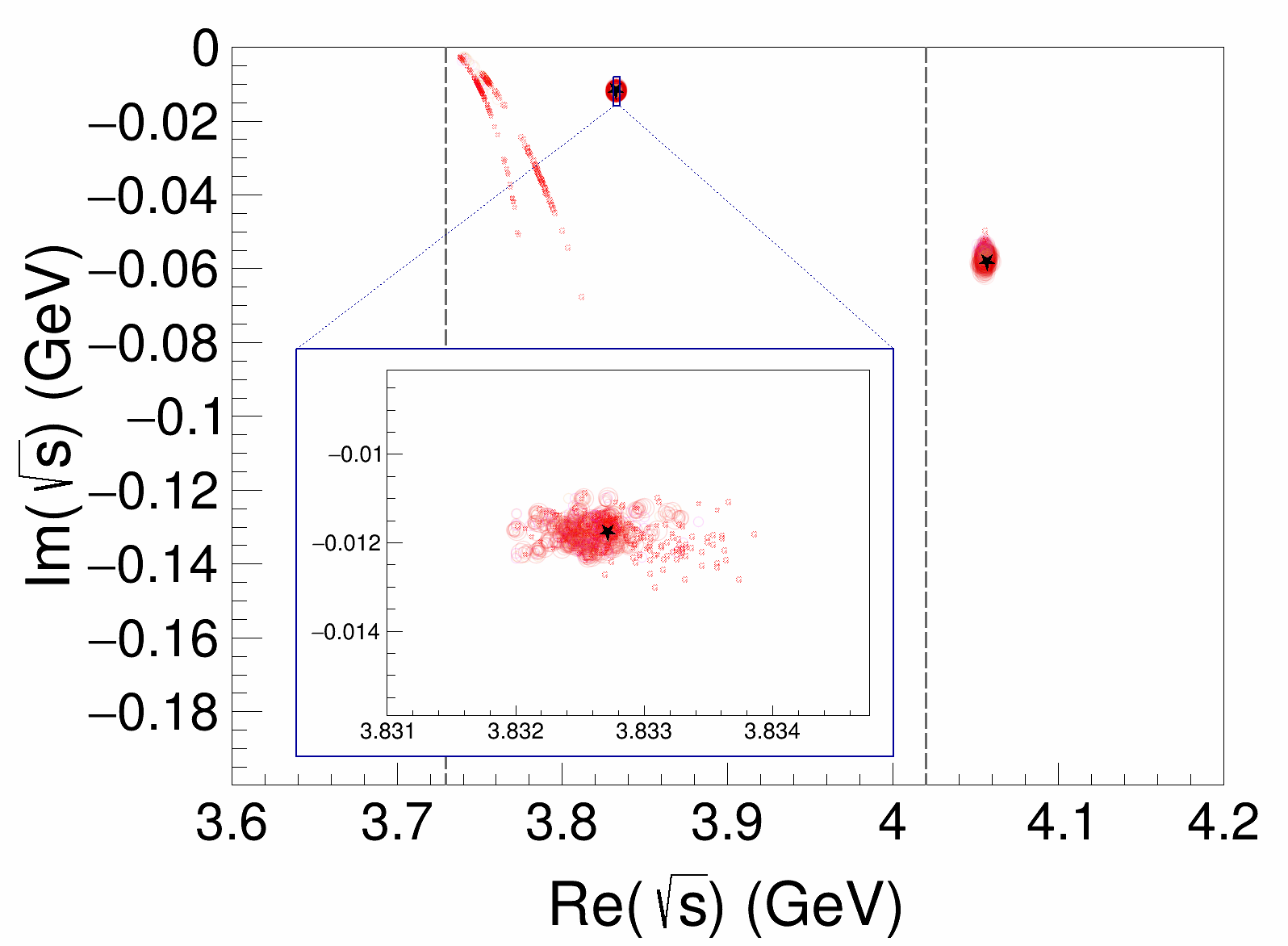}
  \qquad
    \includegraphics[width=0.45\columnwidth]{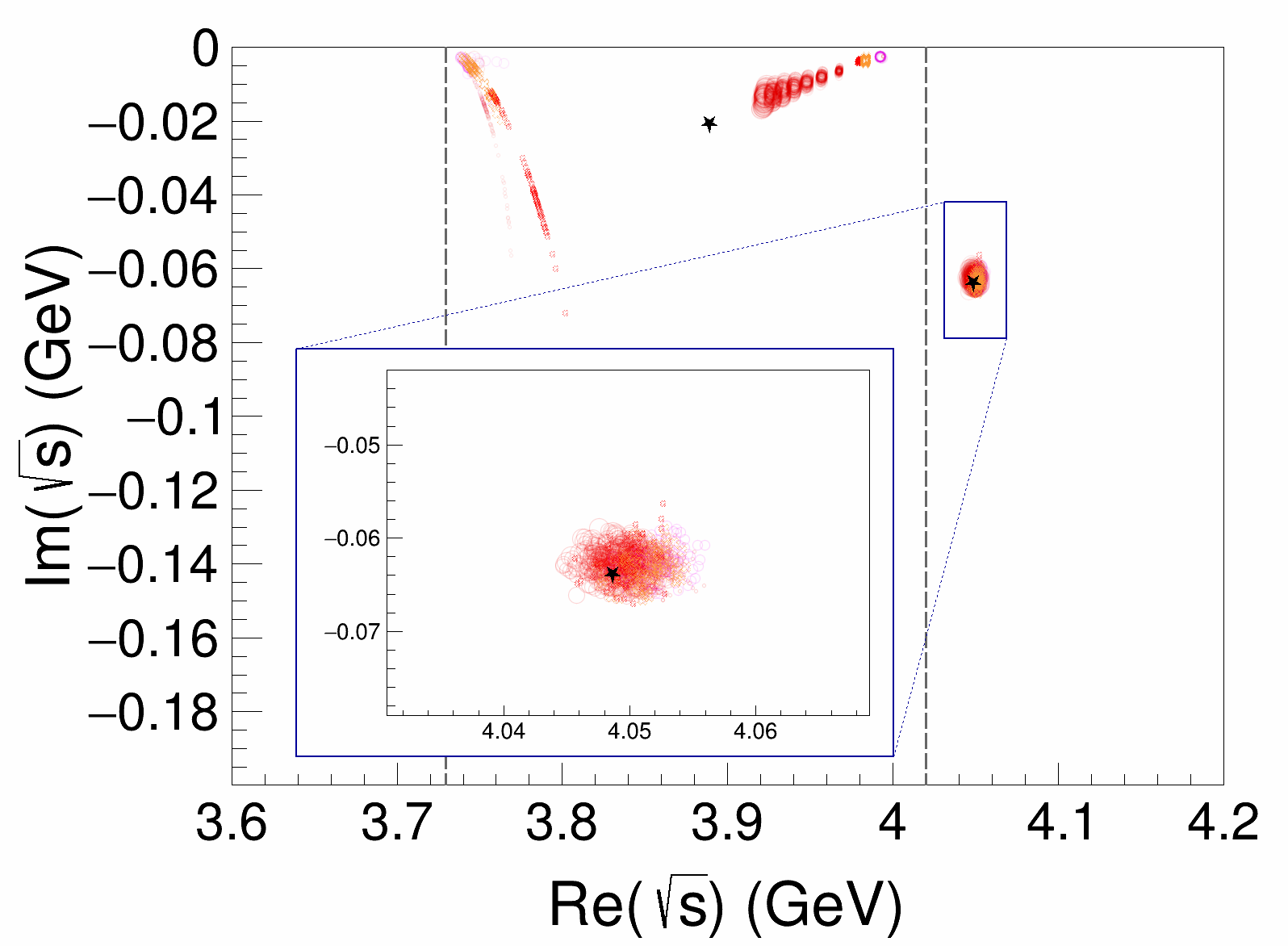}
    \caption{No bare poles, dispersive models with all three form factor choices and second order polynomials in $s$ as non-resonant terms. Sheet $\langle -+\rangle$ (II) (left); sheet $\langle --\rangle$ (III) (right). Symbols are as in Fig. \ref{fig:2pole}.}
  \label{fig:0pole}
\end{figure}

In the final example, we fit the synthetic data with a model that uses a second-order polynomial in $s$ for the background scattering and no resonant interactions at all (while still varying the form factor models).
Once again, the fit quality was very good, and both poles are accurately reproduced, as shown in Fig. \ref{fig:0pole}.

These examples clearly demonstrate that a good fit to data says very little about the interpretation of the amplitude model. Certainly, it is unwise to assign reality or correctness to amplitude models, just as it is unwise to interpret them or their parameter values.

\subsection{Extraneous Poles }

We have seen that a model without bare poles can describe an amplitude with two physical poles. Since these poles are near the physical region, they are essential for fitting the data and are therefore physically plausible. However, as we have observed, it is not reasonable to interpret them as evidence that states are dynamically generated from hadronic interactions; they are merely features of the amplitude model.

\begin{figure}[ht]
  \centering
  \includegraphics[width=0.5\columnwidth]{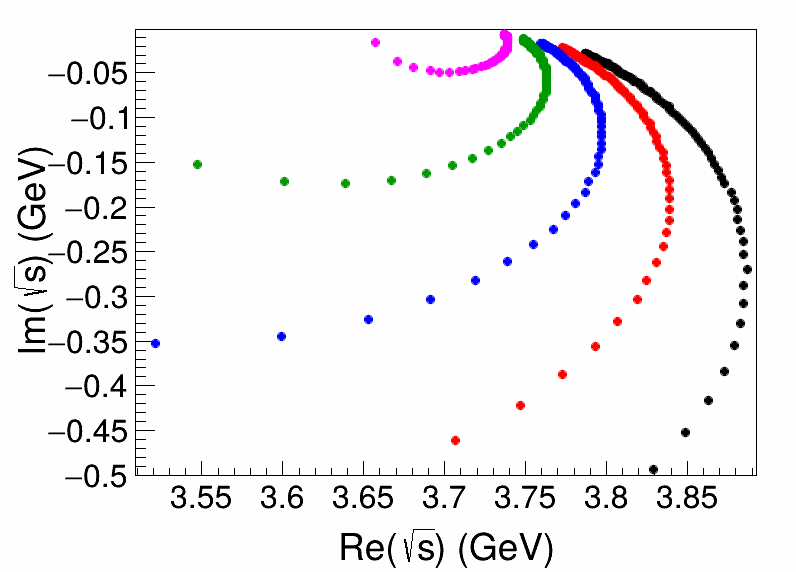}
  \caption{Poles arising from a Blatt-Weisskopf formfactor dispersive K-matrix  model with constant interaction. The value of $\beta$ decreases for different colors from right to left, starting with $\beta=0.2\;\textrm{GeV}$ (pink) up to $\beta=1.0\;\textrm{GeV}$ (black).}
  \label{fig:BlW3}
\end{figure}

To illustrate this, we examine a simple, single-channel dispersive model with P-wave Blatt-Weisskopf form factors and a constant K-matrix. Fig. \ref{fig:BlW3} shows that a pole appears at a position that can be adjusted by changing the form factor scale and the constant. Adding a subtraction constant permits freely moving the pole position along the trajectory shown in Fig.~\ref{fig:BlW3}. Of course, this flexibility can be used to fit data when large amplitude variations are necessary.

Some models produce three poles, and we observe that dispersive models are more prone to generating supernumerary poles. It is tempting to label these poles ``spurious" since they are not required to fit the data; however, in most cases, it is impossible to determine if a pole is spurious using only a single model choice. Comparing a variety of models can lend confidence in conclusions regarding the minimum number of physical poles required to fit a data set. 
For example, our results indicate that models with supernumerary poles form a minority of the approximately 400 models we examined (10-20\% depending on sheet and chi-squared).
Thus, a prudent modeler will test many models and discard results that require more physical poles than the majority of models need. Unfortunately, it is hard to use this as a diagnostic tool because one must thoroughly explore the space of amplitude models to identify the class of models that minimizes the number of physical poles while still fitting the data.

\subsection{Comparison of Dispersive and Nondispersive Models}

We have shown that dispersive and nondispersive K-matrix models can be aligned in the physical region through simple adjustments. However, they generally do not match in the complex plane, which can lead to different pole structures. This issue is explored here by fitting three types of models: nondispersive ($C = -i \rho$), semidispersive ($C = -\int \rho(s')/(s'-s)$), and dispersive ($C = -\int \rho(s') g^2(s')/(s'-s)$). These are paired with the same three form factor models—Blatt-Weisskopf, power, and exponential—using form factor scales from $\beta = 0.2$ GeV to 4 GeV. The same bootstrap method is applied to compare the statistical precision with the systematic differences among models.

\begin{figure}[ht]
  \centering
  \includegraphics[width=0.45\columnwidth]{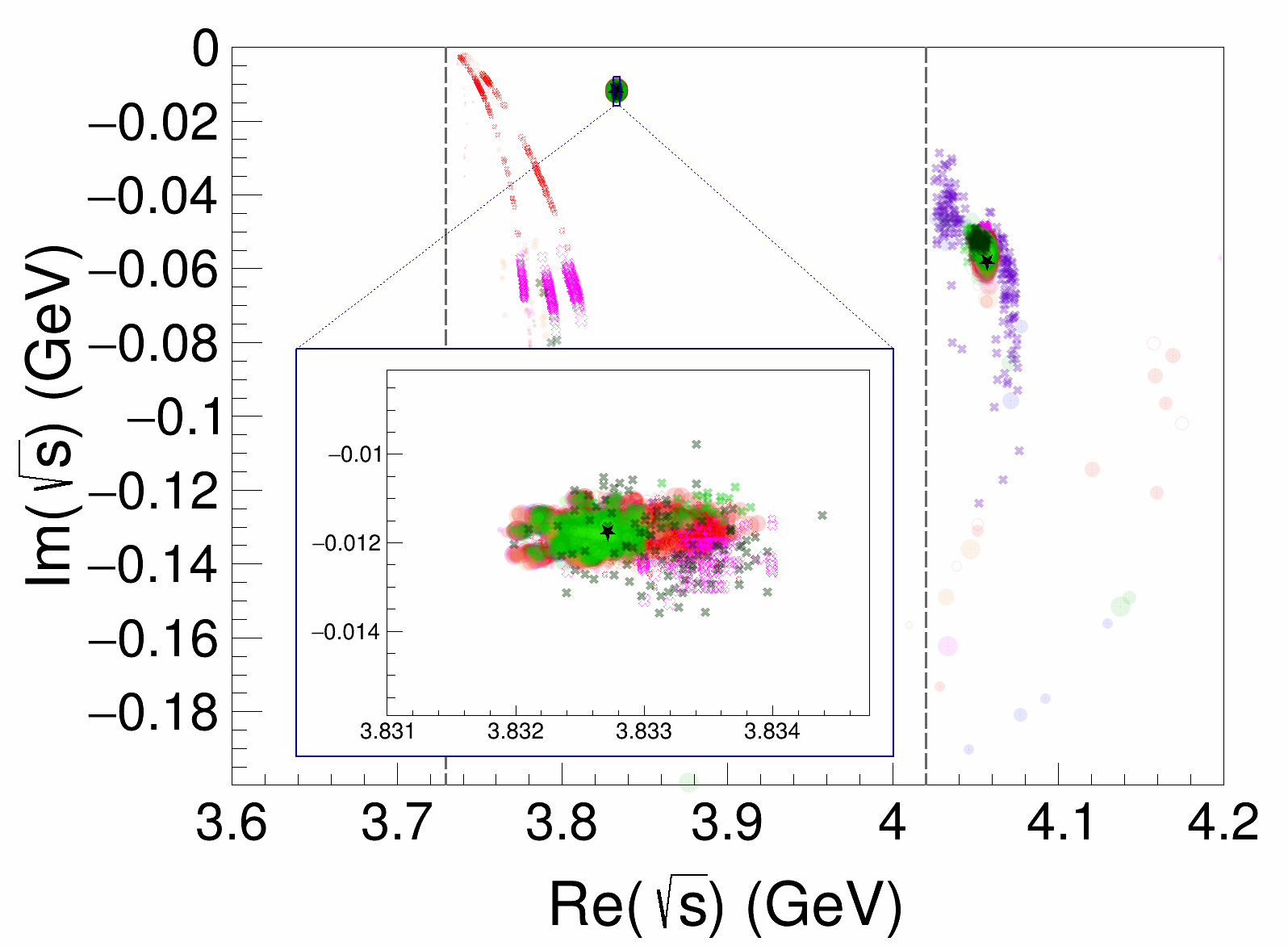}
  \qquad
    \includegraphics[width=0.45\columnwidth]{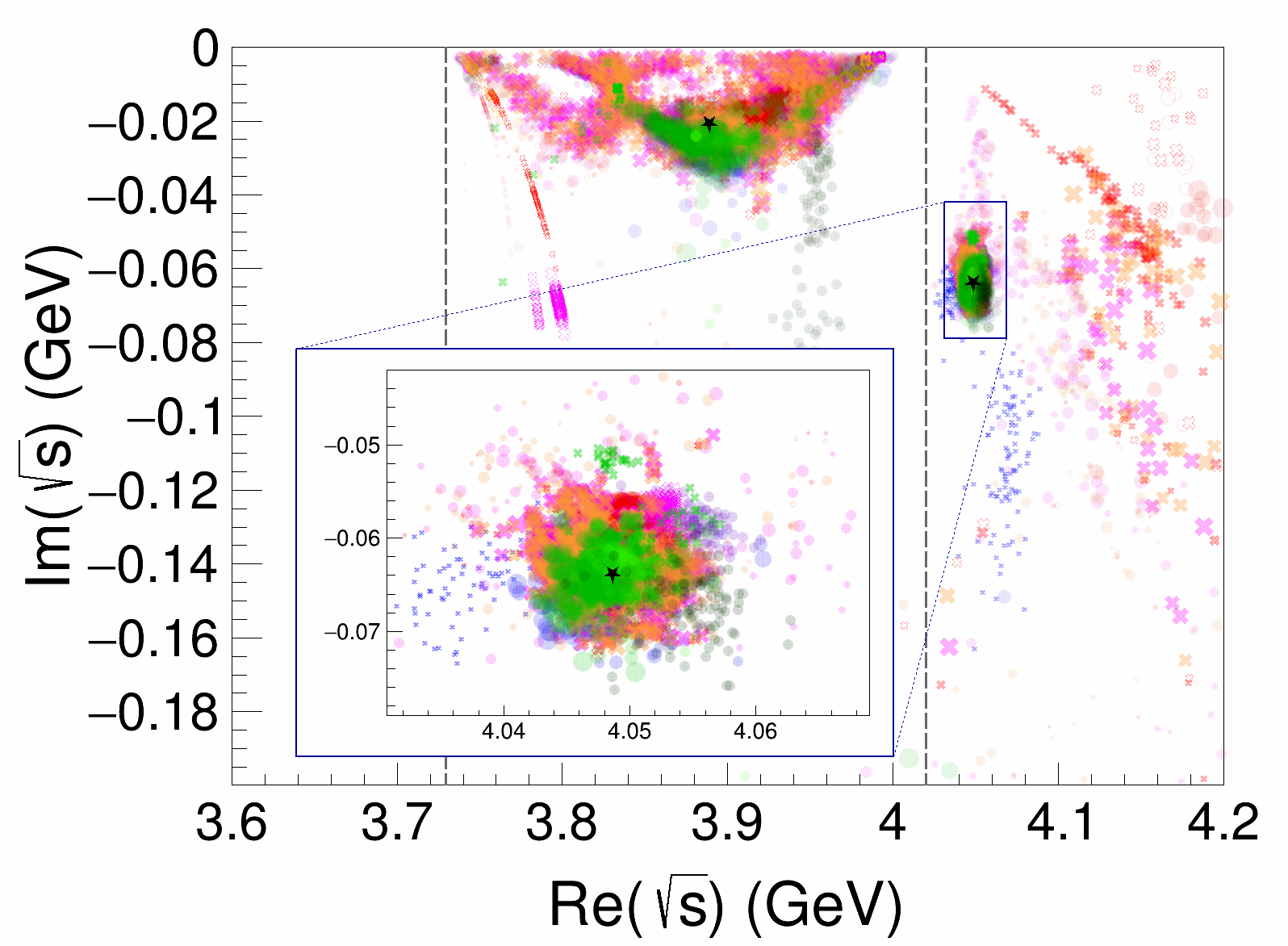}
    \caption{Poles from Approximately 20K Fits. $C=-i\rho$ (green points); 
    semi-dispersive (blue points); other symbols as in Fig. \ref{fig:2pole}.
    Sheet $\langle -+\rangle$ (II) (left); sheet $\langle --\rangle$ (III) (right).}
  \label{fig:all}
\end{figure}

Fig. \ref{fig:all} clearly shows that almost all models accurately reproduce the nominal poles (note that not all model fits are shown). However, on sheet $\langle-+\rangle$ one sees that a series of spurious poles can be produced from dispersive models with a subtraction constant as a free  parameter. These poles often result from fits that produce extraneous poles, indicated by crosses.

A similar, though more random, situation occurs on sheet $\langle--\rangle$. One also sees a large scatter of the lower pole locations, which we interpret as a lack of robustness that becomes more pronounced due to the increased distance from the physical region.

The figure suggests that the most stable among all the amplitude models are those that simply use phase space for the matrix $C$ (green points). The next most stable are the semi-dispersive models, while the fully dispersive models are the least stable. However, fewer results from nondispersive and semi-dispersive models are plotted because these tend to produce more fits that fail to meet the quality cutoff. Overall, all three classes of models perform similarly in locating the nominal poles. Therefore, we conclude that ``less analytic" models perform as well as  ``more analytic" models in terms of fit robustness, which is somewhat counterintuitive. We note, however, that the pseudothreshold lies at $s=0$ in this work, and these conclusions may not hold when the pseudothrehold is closer to the threshold.

\subsection{Residues}

Residues at pole locations reveal the couplings of a resonance to its decay channels. These are defined as

\be
r_{\mu\nu} = -\frac{1}{2\pi i} \oint ds\, \mathcal{M}_{\mu\nu}
\ee
at the resonance pole location $s_\star$. In general, residues are complex-valued and, in the case of narrow widths, are related to the partial width by 

\be
\Gamma(R:\mu) \approx |r_{\mu\mu}|^2 \frac{\rho_\mu(M_R)}{M_R}
\label{eq:res}
\ee
with $M_R = Re(s_\star)$. 

Defining the partial width for broad resonances is more challenging. Perhaps the most straightforward approach is to match to a model amplitude proportional to $g_\mu(s)g_\nu(s)(m^2-s - i \sqrt{s}\sum_\alpha \Gamma_\alpha(s))^{-1}$ and approximate the coupling at the pole location, obtaining

\be
\Gamma(R:\mu) \approx |r_{\mu\mu}|^2 \frac{\rho_\mu(s)}{\sqrt{s}}.
\ee

\begin{figure}[ht]
  \centering
\includegraphics[width=0.65\columnwidth]{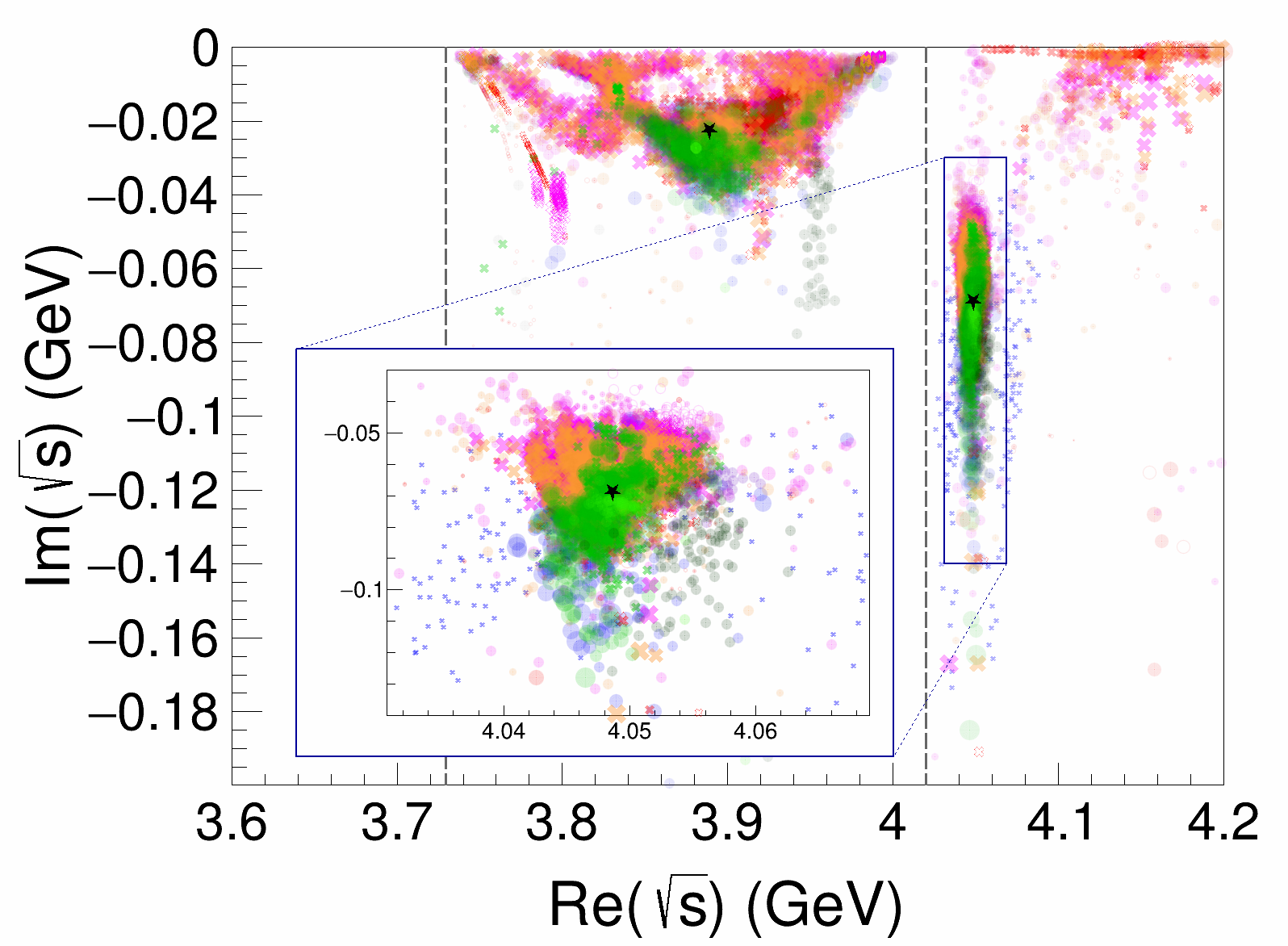}
\caption{Partial widths on sheet $\langle--\rangle$ (III) for the $D\bar D$ channel (all models). Symbols as in Fig. \ref{fig:all}. Stars indicate the nominal residues.}
\label{fig:res}
\end{figure}

We have evaluated the residues by numerically integrating around a small contour centered on the relevant pole location. The resulting partial width (obtained from Eq. \ref{eq:res}) on sheet $\langle--\rangle$ (III) for the light ($D\bar D$) channel is displayed for all models in Fig. \ref{fig:res}.

Comparison with Fig. \ref{fig:all} shows that the residues accurately reproduce the expected partial widths and closely mimic the full widths. The main difference is that the spread in the imaginary part of the higher partial width is significantly larger than in the pole location, likely due to uncertainty in extracting the residues and the narrow width approximation.

\subsection{Additional Checks}
\label{sect:checks}

So far, these conclusions are based on a single set of synthetic data. Here, we examine how robust they are with respect to  variation in the data model. The synthetic data have been assigned Gaussian errors of fixed width, which represent background-dominated errors. An alternative error model suggests that the error varies as the square root of the number of events. In this case,

\be
\Delta \sigma = \sqrt\frac{\sigma}{L\epsilon}
\label{eq:err}
\ee
where $L$ is the experimental luminosity and $\epsilon$ is the detector efficiency. We take $L=200 \text{ pb}^{-1}$ or 2 fb$^{-1}$ per 10 MeV bin, and $\epsilon = 1$\%. These roughly correspond to current and planned specifications of the Belle II detector. The smaller luminosity model produces results nearly identical to those presented above, indicating that the specific error attribution details are not crucial for the considerations here.

It is interesting to analyze how the precision of the data affects our conclusions. Therefore, we fit the entire set of models to synthetic data generated with Eq. \ref{eq:err} for $L=2\text{ fb}^{-1}$. In general, the results are similar to Fig. \ref{fig:all}. As expected, the poles are more tightly clustered around the nominal poles and more models fail to meet the fit criterion. As before, the phase space and form factor models have little influence on the fit quality. Finally, we observe that models with fewer than two bare poles have more difficulty fitting the  data effectively (although using large form factor scales can improve those fits).

\begin{figure}[ht]
  \centering
\includegraphics[width=0.45\columnwidth]{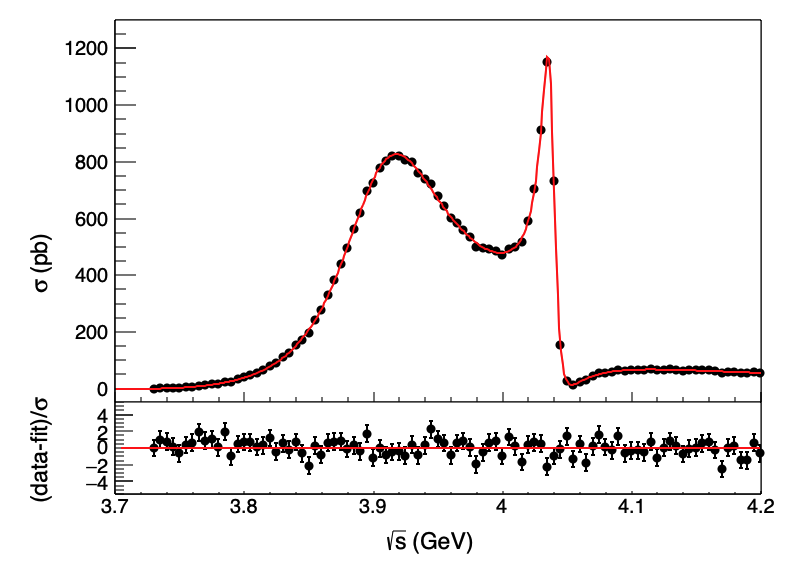}
 \qquad
\includegraphics[width=0.45\columnwidth]{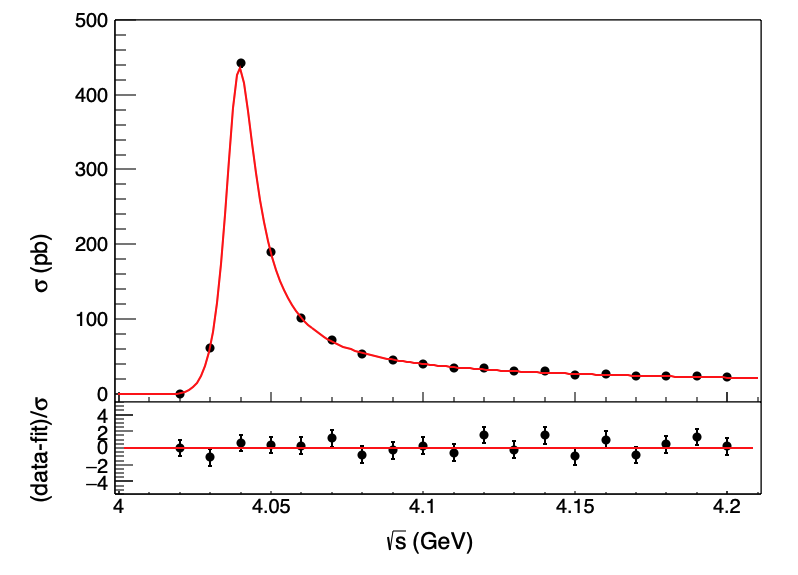} 
    \caption{Alternative model with precise data.}
  \label{fig:modelB}
\end{figure}

\begin{figure}[ht]
  \centering
\includegraphics[width=0.45\columnwidth]{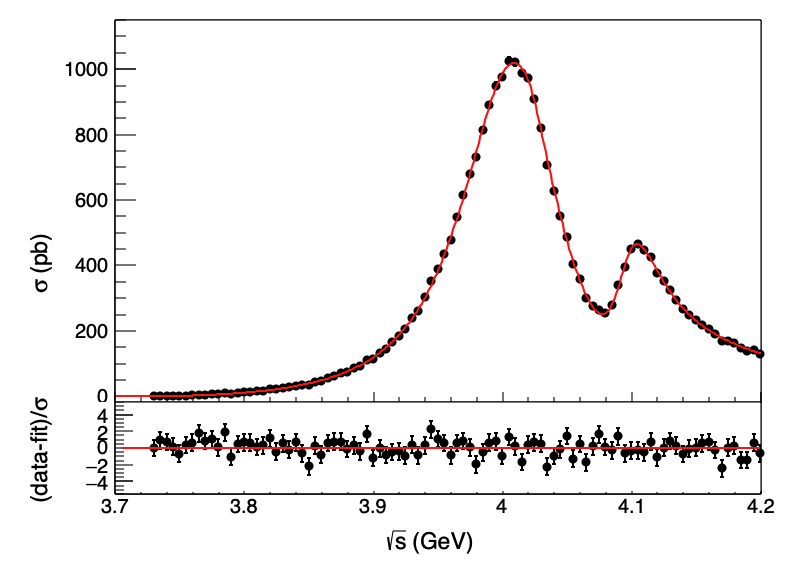}
 \qquad
\includegraphics[width=0.45\columnwidth]{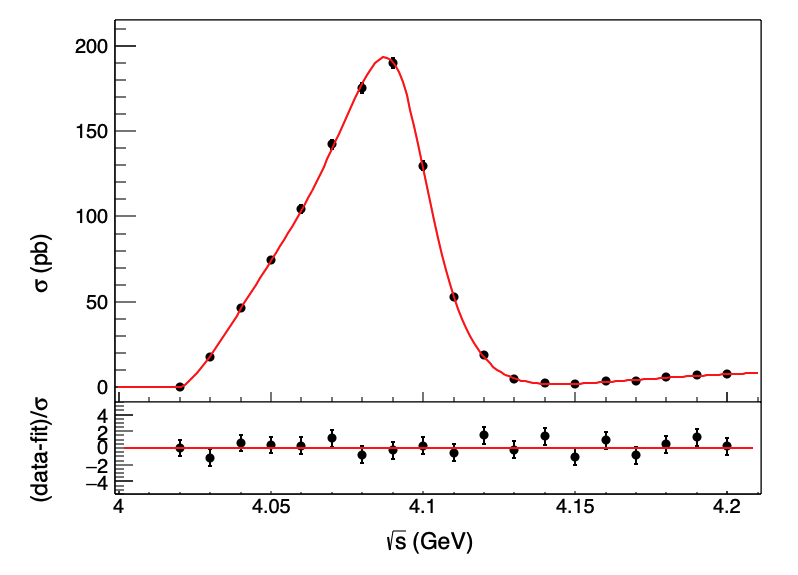} 
    \caption{Alternative model II  with precise data.}
  \label{fig:modelC}
\end{figure}

Finally, we also created two additional synthetic datasets that emphasize channel couplings by placing the second pole near the threshold (Fig. \ref{fig:modelB}) and by creating broad overlapping resonances (Fig. \ref{fig:modelC}).
 The data and a fit are shown for the $L= 2 \text{ fb}^{-1}$ error model in the respective figures. 
We find that the main results of the previous analysis still hold, making it plausible that our conclusions have broad validity.

\subsection{Comment on $G(3900)$}

The evident difficulty in interpreting the  peak at 3900 MeV observed in the BESIII measurement of $e^+e^-\to D\bar D$~\cite{BESIII:2024ths} served as partial motivation for this study.  One group was able to fit the data well with a two-bare pole semi-dispersive K-matrix model that produced two physical poles~\cite{Husken:2024hmi}. However, another group fit the data with a two-bare pole dispersive K-matrix model with subtraction constants that gave rise to three physical poles~\cite{Lin:2024qcq}. The authors of this study then concluded that the third pole corresponds to a dynamically generated physical state. We have observed that creating extraneous poles is  common in dispersive models with subtraction constants, which naturally casts doubt on the conclusions of Ref. \cite{Lin:2024qcq}. More importantly, we have definitively shown that interpreting model parameters, and even the model forms themselves, is ambiguous and ill-advised. Solving this problem will require fitting a large and representative set of models to the data (preferably to a broad and representative data ensemble) to identify the preferred number of physical poles. Meanwhile, the principle of parsimony suggests that models with the fewest physical poles should be favored.

\section{Discussion and Conclusions}

We have explored the link between the N/D formalism and the K-matrix and used this connection to develop dispersive K-matrix models. It was shown that these approaches are equivalent through reparameterization in the physical region. We also presented a general method for continuing between sheets. Several ambiguities in defining a K-matrix model were addressed. For example, we found that three choices of phase space (nondispersive, semi-dispersive, and dispersive) all result in similar fit quality and accuracy. We also emphasize that especially far from the physical region and with dispersive models, it is easy to generate supernumerary poles, and that model parameters and form should not be interpreted as physical. If one insists on interpreting the model, then simple models, which offer parsimony, are likely more robust than complex ones. Model complexity introduces interpretational uncertainty.

It is possible that fitting a single dataset will uncover features specific to that dataset. No dataset is perfect. Therefore, it is better to fit an ensemble of datasets, such as through bootstrapping, to make conclusions more reliable.

In a completely similar sense, no model is infallible. Our results clarify that model conclusions should be tested against a collection of models. Specifically, the presence of extraneous poles can be identified by examining the set of poles generated by iterating over the dataset and model ensembles. Moreover, spurious analytic structures become more apparent on more distant sheets, and this can serve as a diagnostic tool.

Finally, the good fits to all our synthetic datasets by models with zero, one, or two bare poles clearly show that model parameters, or even the models themselves, should never be considered as real or interpreted as physical entities. The only outputs that can be meaningfully interpreted are the poles and residues of the model, or other observables, such as cross sections. We also demonstrated that these conclusions hold true across a variety of synthetic data models, making them likely to be generally valid.

Several interesting issues still remain. We have addressed methods to handle singularities caused by right-hand and left-hand cuts. Other singularities, such as those arising from triangle diagrams, may be relevant in describing certain reactions. Efficiently incorporating these into amplitude models remains an open challenge. Similarly, modeling reactions with broad particles is mostly unexplored. In these cases, it is possible to approximate the $C$-matrix using an analog of a loop diagram with a broad virtual particle \cite{Basdevant:1978tx}. Of course, a complete treatment requires dealing with three-particle intermediate states. A formalism that allows a practical implementation  in fitting routines remains to be developed. 

Practicality issues also influence unbinned fitting, an increasingly common process that demands significant computing resources. Implementing our advice on exploring model space is therefore challenging, and new methods likely need to be developed.

Finally, the formal aspects of enhancing robustness through data fitting and model ensembles remain unexplored. It would be valuable to investigate how to efficiently sample the model space and quantify the results of this sampling.

\begin{acknowledgments}

The authors acknowledge support from the Helmholtz-Institut Mainz, Section SPECF and the German Research Foundation DFG under Contract No. FOR5327 (H\"usken), 
the U.S.\ Department of Energy under grants No.\ 
DE-SC0019232  (Swanson), DE-FG02-87ER40365  and 
 DE-AC05-06OR23177, under which Jefferson Science Associates, LLC operates Jefferson Lab (Szczepaniak). This work contributes to the goals of the US DOE ExoHad Topical Collaboration, Contract No. DE-SC0023598.
We thank C. Hanhart for many helpful discussions on this topic.

\end{acknowledgments}

\bibliographystyle{apsrev4-2}

\bibliography{disp}

\end{document}